\documentclass[12pt]{article}
\usepackage[dvips]{graphicx}

\makeatletter

\newcommand{\singlespacing}{\let\CS=\@currsize\renewcommand{\baselinestretch}{1}\tiny\CS}
\newcommand{\doublespacing}{\let\CS=\@currsize\renewcommand{\baselinestretch}{1.5}\tiny\CS}

\def\@citex[#1]#2{\if@filesw\immediate\write\@auxout{\string\citation{#2}}\fi
  \def\@citea{}\@cite{\@for\@citeb:=#2\do
    {\@citea\def\@citea{,\linebreak[0]\hskip0pt plus .2em}%
      \@ifundefined{b@\@citeb}%
        {{\bf ?}\@warning{Citation `\@citeb' on page \thepage\space undefined}}%
      \hbox{\csname b@\@citeb\endcsname}}}{#1}}

\makeatother

\usepackage{graphics,graphicx}

\begin{document}

\title{Synchronization in Electrically Coupled Neural Networks}

\author{Rajesh G Kavasseri\thanks{Rajesh G Kavasseri is with the Department of Electrical and Computer
Engineering at North Dakota State University, Fargo, ND 58105 -
5285, USA ~(email: rajesh.kavasseri@ndsu.nodak.edu)}}

\author{Rajesh G. Kavasseri, \\ Department of Electrical and Computer
Engineering \\ North Dakota State University, Fargo, ND 58105 -
5285 \\ ~(email: rajesh.kavasseri@ndsu.nodak.edu) \\   \\
Radhakrishnan Nagarajan  \\ Institute of Aging, University of
Arkansas for Medical Sciences\\ 629 Jack Stephens Drive, Little
Rock, AR 72205
 }

\date{}
\maketitle

\begin{abstract}
\noindent In this report, we investigate the synchronization of
temporal activity in an electrically coupled  neural network model.
The electrical coupling is established by homotypic static
gap-junctions (Connexin 43). Two distinct network topologies,
namely: {\em sparse random network, (SRN)} and {\em fully connected
network, (FCN)} are used to establish the connectivity. The strength
of connectivity in the FCN is governed by the {\em mean gap
junctional conductance} ($\mu$). In the case of the SRN, the overall
strength of connectivity is governed by the {\em density of
connections} ($\delta$) and the connection strength between two
neurons ($S_0$). The synchronization of the network with increasing
gap junctional strength and varying population sizes is
investigated. It was observed that the network {\em abruptly} makes
a transition from a weakly synchronized to a well synchronized
regime when ($\delta$) or ($\mu$) exceeds a critical value. It was
also observed that the ($\delta$, $\mu$) values used to achieve
synchronization decreases with increasing network size.

\end{abstract}


\newpage

\section{Introduction}
Computational models of  neural networks have been found to be
useful in characterizing and validating hypotheses about how
information processing occurs in real nervous systems. For example,
a pulse coupled  neural network (PCNN) model in \cite{izhi_1} is
capable of replicating temporal neural activity such as spindle
waves, sleep oscillations and sustained spike synchrony. However, an
important issue that affects the study of network dynamics is the
choice of  neural coupling, \cite{izhi_1}. Neural coupling is
accomplished by synapses which can be broadly classified in to (a)
chemical synapses and (b) electrical synapses. While several studies
consider the former type of coupling to be the preponderant way of
intercellular communication, recent research has provided increasing
molecular and functional evidence of the latter. More importantly,
\cite{bennett2000} and \cite{galarreta} suggest that neurons could
also use electrical synapses to achieve intercellular communication.
There have also been reports that emphasize the importance of
electrical synapses in the temporal coordination of neuronal
activity, \cite{metzer}, the generation of high frequency
oscillations, \cite{dragu} and the generation of oscillatory
activity, \cite{chrobak}.\\

\noindent  The primary focus of this brief communication is to
quantify the extent of synchronization in two general network models
of electrically coupled neurons. The electrical coupling is achieved
with the help of static homotypic gap-junctions (connexin-43). The
choice of connexin 43 was based on a study \cite{pnas} that
presented molecular evidence for its presence in electrical
connections between pairs of neurons in the visual cortex and
hippocampal regions of the juvenile rat brain. However, the methods
discussed are generic and can be extended to other static and
dynamic gap-junctions. Two distinct network topologies namely fully
connected network (FCN) and sparse random network (SRN) are used to
establish the connectivity between the neurons. For example, Fig.1
illustrates a typical gap-junctional connection between a pair of
neurons. In FCN, the gap-junctions are assumed to be exponentially
distributed with mean conductance ($\mu$). The exponential
distribution was chosen as a possible means to capture the
non-uniform distribution of gap-junctions in neuronal populations.
In the case of FCN, the gap-junction strength between every pair of
neurons is non-zero. An alternate approach to accomplish the
distribution of gap-junctions is to assume a sparse-random network
with a specified density of connections ($\delta$). Unlike the FCN,
the coupling strength between any two pairs in SRN is either zero or
one. SRN can be considered as a special case of FCN, where the
gap-junctional conductance between a pair of neurons greater than a
specified threshold is set to one and those lesser than the
threshold are set to zero. Thus it might
not be surprising to view SRN as a quantized version of the FCN. \\

\noindent While neurons are capable of exhibiting a rich set of
firing patterns, we consider a population of {\em bursting} neurons
in this study. Bursting behavior in neurons is considered important
because bursts increase the reliability of synaptic transmission and
provide a mechanism for selective communication between neurons,
\cite{izhi_reso}. Planar bursters can be classified based on the
bifurcation mechanism that leads to the corresponding burst activity
(see \cite{izhi_review} for a summary). In this  study, we restrict
all the bursters in the population to be of the ``square-wave" or
``fold-homoclinic" type. Burst synchronization in general, consists
of two components (a) synchronization of spikes within a burst, and
(b) synchronization between bursts, \cite{izhi_burst_sync}. In order
to minimize the contribution of the former, we studied the envelope
$v^{env}_i, i = 1 \dots N$ of the bursts obtained by filtering the
membrane potentials $v_i, i = 1 \dots N$. A representative burst and
its corresponding envelope
which approximates the duration of the burst is shown in Fig. 2.\\

\noindent In the present study, we show that increasing mean
conductance ($\mu$) and density of connections $(\delta)$ in the FCN
and SRN result in increased synchronization as reflected by the
synchronization index ($M$) (Sec. 3). The synchronization index
($M$) estimated on the original and their envelopes is discussed in
(Sec.4). It is also shown that magnitude of the gap-junction
strength $(\delta, \mu)$ to achieve increased synchronization in FCN
and SRN decreases with increasing population size.

\section{The Model}
In this study, we represent a network by three attributes, namely
(a) individual neurons, (b) the connection between neurons and (c)
the pattern of connectivity.

\subsection{Individual Neuron : Model}
While several models are available to represent a single neuron,
we choose the recently proposed model by Izhikevich \cite{izhi_1},
\cite{izhi_2}, \cite{izhi_3} which is described by a set of two
coupled ordinary differential equations
\begin{eqnarray} \dot{v} = 0.04v^2 + 5v + 140 - u + I \label{neu1}\\
\dot{u} = a(bv - u) \label{neu2}
\end{eqnarray}

\noindent with auxiliary after-spike resetting given by

\begin{equation}
\mbox{if $v = +30$mV, then} \\ \left \{ \begin{array}{l}
v \leftarrow c\\
u \leftarrow u + d \end{array}
                 \right.
                 \label{auxreset}
\end{equation}

\noindent In Eqn.(\ref{auxreset}), the variable $v$ represents the
membrane potential of the neuron while $u$ represents a recovery
variable which accounts for the activation of $K^{+}$ ionic currents
and inactivation of Na$^{+}$ ionic currents, \cite{izhi_2}. The
strength of the model is that it can exhibit firing patterns of
every known type of cortical neuron for various choices of the
parameters $a,b,c$ and $d$, \cite{izhi_2}. Moreover, the model is
computationally superior to several other neuronal models, while
being biologically plausible which makes it attractive for
conducting large scale simulations and hence its choice in the
present study. For a complete summary of the neuro-computational
properties of this and other neuronal models, refer to
\cite{izhi_2}.

\subsection{Connections between neurons: Gap Junctions}

\noindent Electrical synapses between neurons unlike chemical
synapses, are fast and play a crucial role in the synchronization of
neuronal activity. A gap-junction link consists of adjacent
hemi-channels called connexons from neighboring cells. Each connexon
being composed of proteins called connexins, whose conformational
structure dictates their conductance. These junctions can be broadly
classified into homotypic and heterotypic junctions. The variation
of the junctional conductance with the transjunctional voltage is
symmetric for homotypic junctions which is attributed to identical
connexins. However, an asymmetric variation is characteristic of
heterotypic junctions. In this study, we implicitly assume the
junctions to be represented by the homotypic Connexin 43 (Cx43-Cx43)
which has been suggested to mediate neural communication,
\cite{pnas}. However, the methods to be discussed are generic and
can be extended to other types of gap-junctions. Several models have
been proposed in the past to model the dynamics of gap-junctions
\cite{harris}, \cite{vogel}, \cite{chenizu}. In the present study,
we choose the contingent gating model \cite{chenizu}, where the
conductance of a channel between adjacent cells is given by,

\begin{eqnarray}
g_n(v) = g_{res} + P_o(g_{max} - g_{res}) \label{gj1} \\
P_o = \{ 1 + e^{a_0(-v - v_1)} + e^{b_0(v - v_2)} \}^{-1}
\label{gj2}
\end{eqnarray}

\noindent In the above expression, $g_{res}$ represents the
normalized residual conductance of the gap-junction, $g_{max}$ the
normalized maximum conductance, $P_o$ the open probability of the
gap-junction; $\;a_0,\;b_0\;$ are the voltage sensitivity
coefficients; $v_1,\;v_2$ represent the voltages for half-maximal
conductance. In Eqn.(\ref{gj2}), $v$ represents the transjunctional
voltage which is the difference between the membrane potentials of
the two cells connected by the gap-junction.  Considering a
homotypic channel (Cx43-Cx43)with identical gates (as suggested in
\cite{pnas}), the coefficients in Eqn.(\ref{gj2}) have to satisfy
$a_0 = b_0$ and $v_1 = v_2$, \cite{chenizu}. In this case, the
variation of junctional conductance is symmetric with the
transjunctional voltage which means $g_n(v) = g_n(-v)$. Further, the
gap-junctional conductances are assumed to be static. This
implicitly assumes the duration of the action potential to be much
smaller than the changes in the gap-junctional conductance. The
low-dimensional nature of the static contingent gating model makes
it ideal for the present study, hence its choice. The values of the
parameters in Eqn.(\ref{gj2}) for the choice of homotypic connexon
pairing (Cx43-Cx43) are presented in Table 1, \cite{chenizu}. With
the present model, a different connexon pairing can be obtained by
choosing the appropriate values of the parameters ($g_{res},
g_{max}, a_0, v_1, b_0 $ and $v_2$) for that pairing from Table 1,
\cite{chenizu}. While several factors affect the strength of
gap-junctions, in the present study we represent the cumulative
effect by a single parameter ($S$). Thus, the conductance of a gap-
junctional channel between two neurons ($i,j$) in the network is
given by
\begin{equation}
g_{ij} = S \times g_n(v),\;\; v = (v_i - v_j)
\end{equation}
\noindent A schematic representation of a pair of neurons $(i,j)$
connected by a symmetric  homotypic gap junction is shown in Fig.
1.

\subsection{Pattern of Connectivity : SRN and FCN}  In an ideal network,
the neurons should be connected to accurately reflect the pattern of
neural interconnections in the real brain. However, a lack of
precise knowledge of the anatomical connectivity makes such a
construction difficult in general. Here, we consider two models for
the pattern of interconnections namely (i) a sparse random network
(SRN) with fixed conductances through out the network and (ii) a
fully connected network (FCN) where the conductances are
exponentially distributed about a mean conductance $\mu$. In the
sparse random network model, we consider the gap-junctional strength
$(S)$ to be fixed between any two neurons, i.e. $S_{ij} = S_{ji} =
S_0$ if the pair of neurons ($i,j$) are connected. The random
network is created by generating a symmetric matrix $W$ ($ N \times
N$) with density $\delta$ such that $W$ has $(N^2 \delta)$ non-zero
entries. Here, the density $\delta$ of the SRN is defined as the
($\#$ of connections)/$N^2$. Then, a connection between a pair of
neurons ($i,j$) is assigned only if $W(i,j) \ne 0$. If $W(i,j) = 0$,
then no connection exists between neuron $i$ and neuron $j$. Thus,
the sparsity of the network is controlled by the parameter $\delta$.
In FCN, we assume a fully connected structure, that is, every pair
of neurons in the network is connected. However, the connection
strengths $S_{ij}$ are assumed to be exponentially distributed about
a mean strength $\mu$. In both networks, note that the connections
are made symmetric to enforce the bi-directional nature of
gap-junctional couplings.

\subsection{Overall Network Model}
The overall network model is obtained by combining
Eqns.(\ref{neu1}, \ref{neu2}) for the neurons,
Eqns.(\ref{gj1},\ref{gj2}) for the gap junctional conductances and
summing the contributions of synaptic current injections resulting
from the appropriate network topology. In complex physiological
systems such as neural assemblies, discrepancies in ionic
exchanges between the neurons and their environment render the
quantitative dynamics of individual neurons different from one
another.  Therefore, the neural population is modeled as set of
$N$ bursting neurons whose intrinsic parameters $c_i,d_i$ and the
thalamic input $I_i$ are perturbed so that each neuron has a
different burst duration and frequency. This is done by setting
$a_i = 0.02, b_i = 0.2$ and $(c_i, d_i) = (-65,8) + (15, -6)r_i$
where $r_i$ is a random variable uniformly distributed in [0,1].
Thus, the dynamics of the $i$ th neuron in the network is
described by

\begin{eqnarray} \dot{v_i} = 0.04v^2_i + 5v_i + 140 - u_i + I_i + \sum_{j} g_{ij}(v_j - v_i)\\
\dot{u_i} = a_i(b_i v_i - u_i)
\end{eqnarray}

\noindent with auxiliary spike-resetting Eqn.(\ref{auxreset})

\section{Measures of Synchronization}

Several measures have been proposed in the past \cite{phyrevE} to
determine the extent of synchronization in neural systems. These
can be broadly categorized into linear and non-linear measures.
While nonlinear measures may be more appropriate in the present
context, their estimation can be quite challenging. In this
report, we use pair-wise linear correlation ($ 0 \leq \rho \leq
1$) and Morgera's covariance complexity $(0 \leq C \leq 1)$
\cite{morgera} to quantify the extent of synchronization between
the neurons.

\subsection{Pairwise Correlations}
The  index ($\rho_{ij}$) between two neurons ($i,j$) is given by
the linear correlation between their membrane voltages $v_i, v_j$
respectively. For a system of $N$ neurons, we obtain $P =
{}^{N}C_2 = N(N-1)/2$, pair-wise dependencies. Statistically
significant pair-wise dependencies were chosen as those whose
$p$-values are lesser than a specified level of significance
($\alpha = 0.05$). The null hypothesis addressed is that there is
no significant correlation between a given pair of neurons. It
should be noted in the present study we have $P$ pair-wise
dependencies. In statistical literature, the choice of multiple
testing correction such as Bonferroni correction is often
recommended to minimize the false-positive rate. The adjusted
significance is given by $\alpha_{*} = \alpha/P$. For example, in
a ten element ($N = 10)$ network, we have $P = 45$ and therefore,
$\alpha_{*} = \alpha/P = 0.05/45 = 0.0011$. In the present study,
the proportion of significant pair-wise correlations was
determined as those whose p-values were lesser
than $\alpha$ divided by $P$. \\

\noindent To determine whether the distribution of the $(P)$
pair-wise dependencies across any two states of activity for the
network is statistically significant (i.e. $\alpha = 0.05$), we used
parametric (t-test) and non-parametric (Wilcoxon-ranksum) tests.
Unlike the parametric test (t-test), non-parametric tests
(Wilcoxon-ranksum) do not assume a normal distribution of the values
and hence unbiased. However, non-parametric tests are based on ranks
and not the original values, which is a limitation. Thus using a
combination of parametric and non-parametric tests can minimize
spurious conclusions that are an outcome of a particular test's
assumption.

\subsection{Morgera's Covariance Complexity}
\noindent  Morgera's covariance complexity is a linear measure and
has been used recently for the analysis of brain signals
\cite{morgera}, \cite{watanabe}. Here, we use this measure to
define a synchronization index as follows. After simulating the
activity of the network for a given interval ($T$) using a time
step of $h = 0.5 $ ms, the resulting $M = T/h$ discrete time
points of the membrane voltages of the N neurons in the system is
represented as a matrix $\Gamma_{M \times N}$ where ($N << M$).
Singular value decomposition (SVD) of $\Gamma$ yields $N$
eigenvalues which explain the variance along the orthogonal
directions in N-dimensional space. The normalized variance along
the $i$ th component is given by

\begin{equation}
\sigma_i = \frac{\lambda^2_i}{\sum_{i=1}^{i=N}\lambda^2_i}, \;\; i
= 1 \dots N
\end{equation}
Then, Morgera's covariance complexity $C$ is given by,

\begin{equation}
C = -\frac{1}{\log N}\sum_{p=1}^{p=N} \sigma_p \log{\sigma_p}
\end{equation}

\noindent The synchronization index ($M$) is given by $M = (1-C)$
and lies in the closed interval [0,1]. A minimum value of $(M=0)$ is
obtained in the case of random behavior, whereas a maximum value
$(M=1)$ is obtained in the case of perfect synchronization. However,
due to external noise and nonlinear effects, the estimate of $M$
deviates from these extreme values.

\section{Simulation Results}

The effect of increasing gap-junctional conductance on the
synchronization was determined for a given population of neurons.
Two distinct network topologies namely, FCN and SRN were
considered. The synchronization is studied with increasing $\mu$
in case of FCN and increasing $\delta$ in case of the SRN. The
synchronization index (M) was estimated on the actual membrane
potentials ($v_i, i=1 \dots N)$ and their envelopes $(v_i^{env},i=
1 \dots N)$. The choice of the envelopes was in order to minimize
the effect of spikes within bursts on the synchronization index.
In order to determine the effect of varying population size on the
synchronization, we investigated three different populations
namely $(N = 10, 50$ and $100)$. \\

\noindent The estimate of $M$ determined on 20 independent trials
with varying gap-junctional conductance and population sizes is
shown in Figure 3. In the case of FCN, the synchronization index
saturates around $(M \sim 0.9)$ for parameter $(\mu = 40, 20$ and
$5)$ and population sizes $(N = 10, 50$ and $100)$, Figure 3 (top).
An abrupt transition in the synchronization index is observed at a
critical density of $\delta \sim 0.2$ for all three network sizes,
Fig.3 (bottom). To generate a quantized version of the FCN, we
constructed SRN, where the connection strengths $S_0$ was fixed at
40, 20 and 5 for populations $(N = 10, 50$ and $100)$ respectively.
While the connection strength ($S_0$) is fixed between connected
pairs of neurons, their density $\delta$ (Sec 2.2, Eqn.(6)) which
controls the number of connections is gradually increased. It should
be noted that while $\delta \in [0,1]$ is a normalized measure of
network density, the actual number of connections ($N^2 \delta$)
varies in the range $[0 ,\; N^2]$ for a network of size $N$. For
clarity, we show the variation of the synchronization index $(M)$
with respect to ($S_0 \times \delta = S_0 \times \# $number of
connections$/{N^2}$) which we shall denote by $\bar{S}$. Note that
$\bar{S}$ quantifies the average connection strength of the SRN. As
with the case of FCN, we find the network synchronizes with
increasing $\bar{S}$. The synchronization is achieved at a magnitude
similar to FCN. As expected, increasing $(\mu)$ or $\bar{S}$ leads
to increasing synchronization of the network. Interestingly, the
magnitude of the parameter $(\mu, \bar{S} $) required to achieve
synchronization decreases with
increasing population size in the FCN and SRN respectively, Fig. 3. \\

\noindent While the estimate of $M$ obtained on the membrane
potentials and their envelopes converge to a similar value with
increasing $(\bar{S}$ or $\mu)$, the abrupt transition is better
reflected by those estimated on the envelopes, Figure 3. This
behavior was consistent varying population sizes $(N = 10, 50$ and $100)$. \\

\noindent Parametric (t-test) and non-parametric
(Wilcoxon-ranksum) statistical tests were used to determine
statistically significant change ($\alpha = 0.05$) in the
distribution of the pair-wise linear correlation before and after
synchronization. The parameters $(\bar{S} = 0)$ and $(\mu =0)$
were used to represent the networks SRN and FCN before
synchronization. The $(\mu$) and $(\bar{S}$) values after
synchronization for the various population sizes $(N = 10, 50$ and
$100)$ were chosen as $(\mu, \bar{S} = 40, 20$ and  $5)$ from
Figure 3. The distribution of the pair-wise linear correlations
before and after synchronization for the various population sizes
for FCN and SRN is shown in Figure 4. The distributions were
statistically significant after Bonferroni correction. This result
was consistent for all three population sizes.

\section{Conclusions}
In this brief report, we studied the synchronization of temporal
activity in a neural network with electrical synaptic couplings. The
neuronal parameters were randomized to yield different burst
durations and frequencies but constrained to restrict their behavior
to {\em square-wave} type bursting. The electrical couplings were
established by a generic model to represent the homotypic,
bi-directional and static gap-junctional couplings formed by
Connexin-43. Two network topologies namely, (a) sparse random
network (SRN) and (b) a fully connected network (FCN) were used to
model the pattern of connectivity. The overall mean conductance of
the network was governed by $\delta$ the density of interconnections
and $S_0$ the fixed connection strength in SRN, and $\mu$ the mean
conductance in FCN. Synchronization of burst activity was quantified
using (a) the distribution of pairwise linear correlations and (b)
Morgera's covariance complexity. Numerical simulations were
conducted to study the synchronization of the network with
increasing coupling strength and varying population sizes. \\

\noindent We observe that in both networks, an increase in the
mean gap junctional coupling strength ($\bar{S}, \mu$) results in
an increased degree of synchronization. In addition, we observe
that the network exhibits an {\em abrupt} transition from a poor
to highly coordinated regime at a critical value of these
parameters. Further, while the magnitude of the parameter $\mu$
required to achieve synchronization decreases with increasing
population size in FCN, the abrupt transition to increased
synchronization seems to occur at a network density of $\delta
\sim 0.2$ for all three network sizes in case of the SRN. \\

\noindent Our results in general, support the prevailing notion that
gap-junctions promote synchrony in large neural assemblies. The
computational studies show that in addition to coupling strength of
the gap-junctions, the density of their connections in the two
topologies (SRN and FCN) also plays an important role in promoting
synchrony. Moreover, the results suggest the existence of  {\em
critical} parameters at which a large neural network could tune, or
transition to a high degree of synchronization.

\newpage

\begin{figure}[htbp]
\begin{center}
\includegraphics[height=3in,width=4.2in]{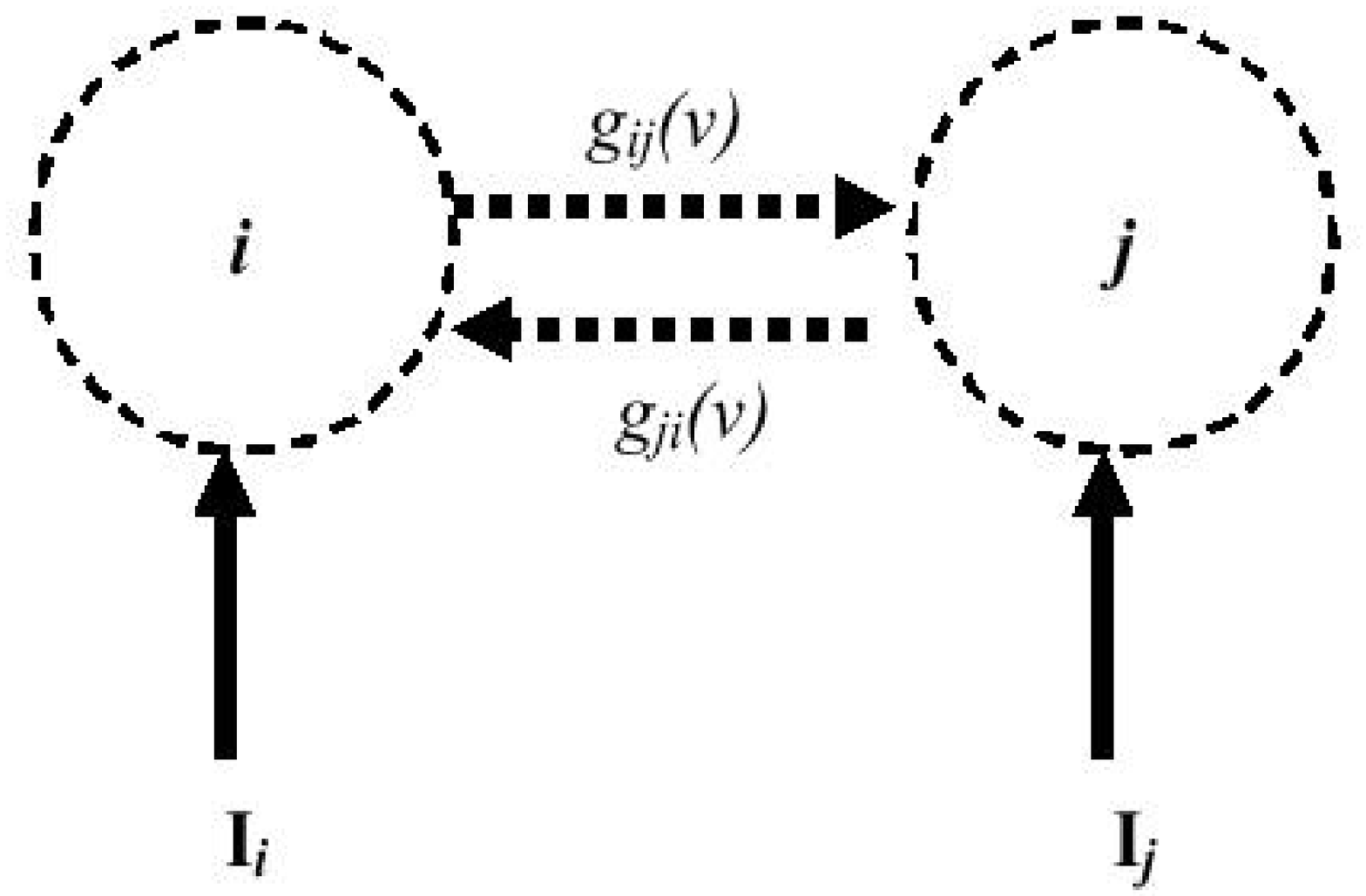}
\caption{A pair of neurons coupled by a symmetric homotypic gap
junction such that $g_{ij}(v_{ij}) = g_{ji}(v_{ji})$ where $v =
v_{ij} = - v_{ji}$ is the transjunctional voltage. The tonic
activation currents are $I_i$ and $I_j$ for neurons $i$ and $j$
respectively.}\label{neurnonij}
\end{center}
\end{figure}

\newpage

\begin{figure}[htbp]
\begin{center}
\includegraphics[height=3in,width=5in]{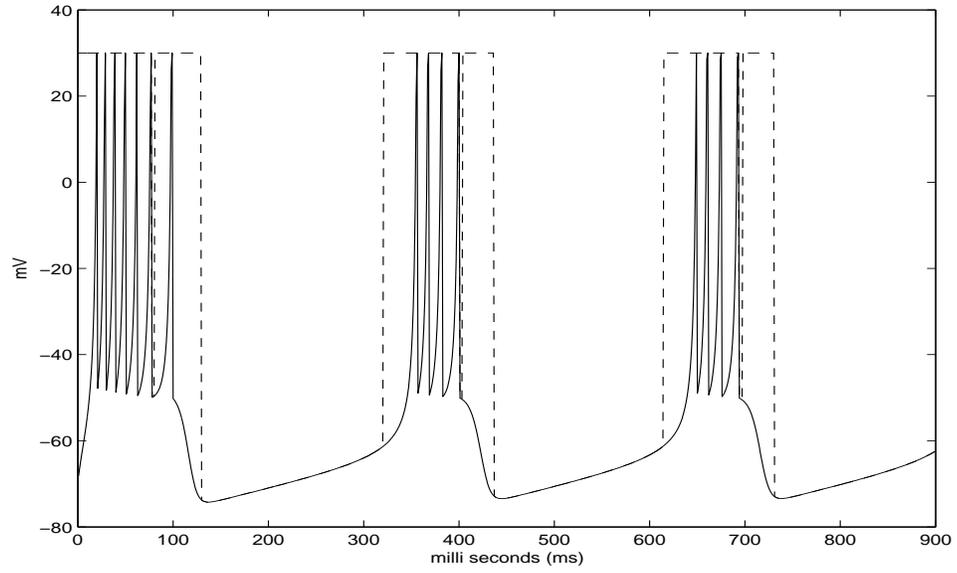}
\caption{A representative burst (solid line) and its corresponding
`envelope' (dotted line). The envelope approximates the duration of
the burst.}\label{neurnonij}
\end{center}
\end{figure}

\newpage

\begin{figure}[htbp]
\begin{center}
\includegraphics[height=5in,width=6in]{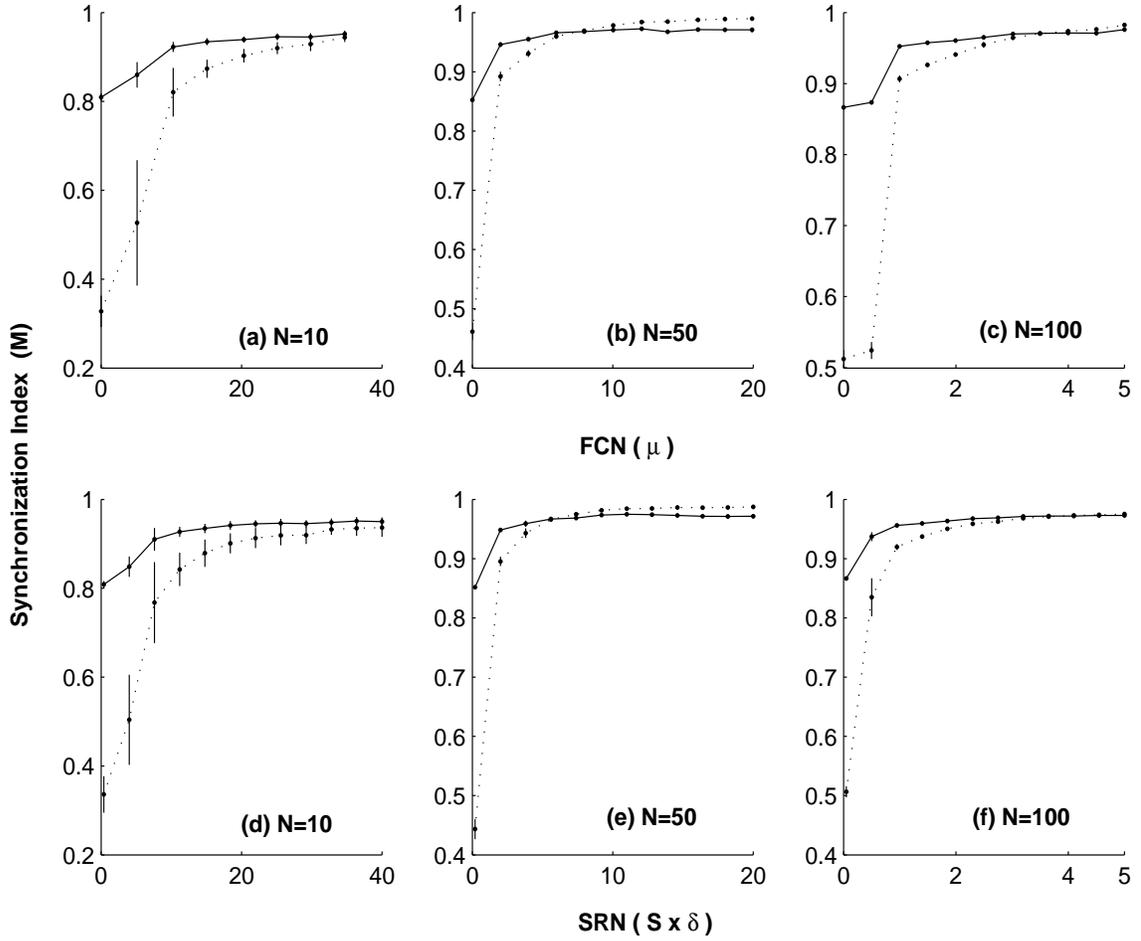}
\caption{Variation of the synchronization index $(M)$ with
parameters $\mu$ and $\delta$ for the fully connected network
(top) and sparse random network (bottom). The vertical lines
represent the variance about the mean value (dots) for twenty
independent trials. Estimates of $(M)$ obtained on the original
waveforms $v_i, i = 1 \dots N$ (solid lines) and the envelopes
$v^{env}_i, i = 1 \dots N$ (dotted lines) for three different
population sizes $N = 10,50$ and 100 is also shown.
}\label{morgpic}
\end{center}
\end{figure}

\newpage

\begin{figure}[htbp]
\begin{center}
\includegraphics[height=5in,width=6in]{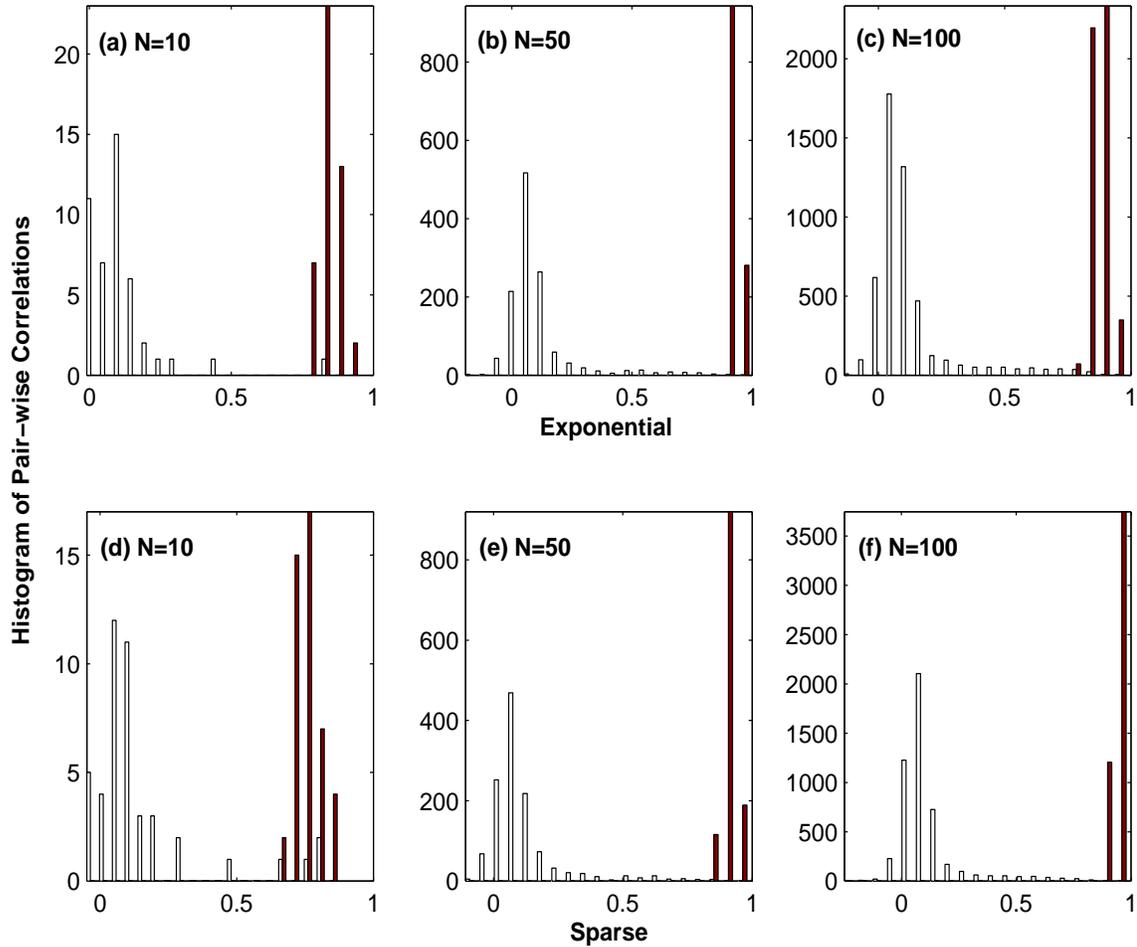}
\caption{Distribution of the pair-wise linear correlations estimated
before and after synchronization for FCN (top) and SRN (bottom).
Pair-wise correlation before synchronization for FCN and SRN were
obtained by setting $(\mu =0)$ and $(\bar{S} = 0)$, represented by
hollow bars. Distribution of the pair-wise correlation after
synchronization for parameters $(\mu = 40, 20$ and 5, top) and
$(\bar{S} = 40, 20$, 5, bottom) for population sizes $(N = 10, 50$,
100) is shown in $(a,b,c)$ and $(d,e,f)$
respectively.}\label{pairwise}
\end{center}
\end{figure}

\end{document}